\documentclass[12pt,epsf]{article}

 \usepackage{epsfig}
 \usepackage{epstopdf}
\usepackage{graphicx}
 \usepackage{times}
  \usepackage{amsmath}
 \usepackage{amssymb}

\newcommand{\be}{\begin{equation}}
\newcommand{\ee}{\end{equation}}
\def\({\left (}
\def\){\right )}
\def\[{\left [}
\def\[{\right ]}

\begin{document}
\begin{titlepage}
\bigskip

\rightline
\bigskip\bigskip\bigskip\bigskip
\centerline {\Large \bf { States of Negative Energy and $AdS_5 \times S_5/Z_k$}}
\bigskip\bigskip
\bigskip\bigskip

\centerline{\large Keith Copsey and Robert B. Mann}
\bigskip\bigskip
\centerline{\em Department of Physics \& Astronomy, University of Waterloo}
\centerline{\em 200 University Avenue West, Waterloo, Ontario N2L 3G1, Canada}
\centerline{\em kcopsey@scimail.uwaterloo.ca, rbmann@sciborg.uwaterloo.ca}
\bigskip\bigskip

\begin{abstract}
We develop a careful definition of energy for nonsupersymmetric warped product asymptotically $AdS_d \times M_q$ solutions which include a nonzero p-form.  In the case of an electric p-form extending along all the AdS directions, and in particular in the case of self-dual fields like those used in the Freund-Rubin construction, the Hamiltonian is well defined only if a particular asymptotic gauge for the p-form is used.  Rather surprisingly, asymptotically this gauge is time dependent, despite the fact the field and metric are not.  We then consider a freely orbifolded $AdS_5 \times S_5$ and demonstrate that the standard boundary conditions allow states of arbitrarily negative energy.  The states consist of time symmetric initial data describing bubbles that are regular up to singularities due to smeared D3-branes.    We discuss the evolution of this data and point out that if the usual boundary conditions are enforced such bubbles may never reach infinity.
\end{abstract}

\end{titlepage}

\baselineskip=16pt
\setcounter{equation}{0}

\section{Introduction}

Orbifolding the SU(4) R-symmetry of $\mathcal{N} = 4$ SYM allows one to break some or all of the supersymmetry of that theory.  One may then study the AdS-CFT conjecture in a context of reduced or no supersymmetry \cite{KachruSilverstein, LawrenceNekrasovVafa, OzTerning}.  In the case of completely broken supersymmetry there is the possibility that the theories on one or both sides of the duality might be unstable.   For an orbifold with fixed points, both sides of the duality have been shown to suffer from instabilities in the winding sector \cite{AdamsSilverstein, APS, Dymaskyetal, ArmoniLopezUranga}.   The gravitational instability corresponds to the existence of closed string tachyons.   In the case of freely acting orbifolds, on the other hand, the twisted sector strings which are tachyonic in the case of orbifolds with fixed points have lengths of order on AdS length scale and acquire a positive mass.   It has been argued, however, that a nonperturbative instability should still be expected \cite{FabingerHorava,AdamsSilverstein, APS}.

Recently Horowitz, Orgera, and Polchinksi have described an instanton describing a ``bubble of nothing'' type instability of a freely orbifolded $AdS_5 \times S_5/{Z_{k}}$ \cite{HOP}.   Like the original Kaluza-Klein bubble of nothing \cite{Wittenbubble}, these solutions describe the creation of a topologically nontrivial spacetime from the vacuum.  The bubble then expands outwards, consuming the space.   These recent solutions differ qualitatively from previous bubbles of nothing studied in AdS.  Most of these examples (\cite{BR} - \cite{HR}) involve modifying the conformal metric on the boundary so that it becomes $S^1 \times dS_3$ while the solutions of \cite{HOP} involve identifications only along the $S_5$, not along the $AdS_5$ portion of the space.  The only other currently known dynamical AdS bubbles \cite{CopseyAdSbubbles} do not involve any identifications, although in that case it is not entirely clear whether or not the solutions describe an instability.

The authors of \cite{HOP} find this instability only when they impose antiperiodic boundary conditions for fermions, as well as D-brane singularities in the most straightforward case, so none of the presently known positive energy theorems apply.  Any solutions in the class of \cite{HOP} with the desired boundary conditions are massless; this is a necessary consequence of their description as a decay of the vacuum.  These solutions include large bubbles which accelerate outward and hence include a substantial amount of energy in terms of gravitational radiation, suggesting there exist bubbles with negative energy.  It is also worth noting that in the asymptotically flat Kaluza-Klein context, in addition to Witten's massless bubble, there are regular arbitrarily negative energy bubbles \cite{BrillPfister,BrillHorowitz}.  With the above considerations one suspects, as do the authors of \cite{HOP},  that spacetimes which are asymptotically $AdS_5 \times S_5/{Z_k}$ admit negative energy solutions.  

We show here that there are regular (up to singularities due to smeared D3-branes) bubbles of arbitrarily negative energy for these boundary conditions. To make this claim, we must properly define energies of nonsupersymmetric solutions of warped product spaces (in particular those asymptotic to $AdS_5 \times S_5/Z_k$) with a p-form flux.  A Hamiltonian definition is developed in section two.  We also point out there that the asymptotic choice of gauge has physical consequences.  In particular, for spaces such as $AdS_5 \times S_5$ with a Freund-Rubin \cite{FR} compactification utilizing a self-dual five form, there is only a single choice of gauge that yields a well defined Hamiltonian.  Rather surprisingly, this choice dictates the potential must be asymptotically time dependent, despite the fact that the fields and metric are not.

With a proper definition of energy in hand, in the third section we consider a large class of time symmetric initial data generalizing the form of solutions considered by \cite{HOP}.  The fourth section describes some particular negative energy solutions, and  the fifth section discusses the time evolution of this data. We point out that while bubbles such as ours and those of \cite{HOP}  may become large, as long as the standard boundary conditions are preserved they never reach infinity.  We conclude with a discussion of some open problems for the gravitational theory and with regards to the AdS-CFT conjecture.

 \setcounter{equation}{0}
\section{Regarding Energy}

We will use the Hamiltonian to define the energy.   There are, of course, many other possible definitions of the energy (see, e.g.,  \cite{HIM, TaylorSkenderis}) but we take the perspective that any other sensible definition must be equivalent to this one.  Beginning with the action for a p-form and scalar in D dimensions:
\be \label{actn}
S = \beta \int d^D x \sqrt{-g} \Big(R - \frac{1}{2} (\nabla \phi)^2 - \frac{e^{-\alpha \phi}}{2 (p + 1)!} H_{\alpha_1 \ldots \alpha_{p + 1}} H^{\alpha_1 \ldots \alpha_{p + 1}} - V(\phi) \Big)
\ee
where $H = dB$ is a (p + 1)-form field strength, $\phi$ is a 
scalar, $\alpha$ is the dilaton coupling, and $\beta$ is a
normalization constant we choose to leave arbitrary.  While we have only labeled a single scalar and p-form, the generalization of the results of this section to multiple fields is entirely straightforward.  Our interest in the later sections of this paper is in solutions where $\phi$ is the dilaton and may be consistently set to zero but  the considerations in this section are of somewhat broader interest and it will not cause us difficulty to include a nonzero scalar. 

Consider the standard Hamiltonian decomposition with the spacelike slice $\Sigma$ with unit timelike normal $n^{\mu}$ and time evolution vector $\xi$.  We will use latin indices to indicate when sums are only over directions in $\Sigma$ (i.e. spatial indices).  The lapse $N$ is given by
 \be
 N = -n_{\mu} \xi^{\mu}
 \ee
 while the shift vector $N^a$ is
 \be
 N^a = h^{a}_{b} \xi^b
 \ee
 where $h_{a b} = g_{a b} + n_a n_b$ is the induced spatial metric on $\Sigma$.

We will denote the Lie derivative of a
     tensor in the $\xi$ direction projected into the surface $\Sigma$ (the ``time derivative'') by a dot:
     \begin{equation} \label{Bdot}
	 {\dot{{B}}_{a_1 \ldots a_n}} = h_{a_1}^{b_1} \ldots h_{a_n}^{b_n} {\mathcal{L}_{\xi} B}_{b_1 \ldots b_n}
	 \end{equation}
 The momentum canonically conjugate to the spatial metric ${h}_{ab}$
is, as usual,
\begin{equation}\label{piG}
\pi_{G}^{ab} = \frac{\partial \mathcal{L}}{\partial
\dot{{h}}_{ab}} = \beta \sqrt{{h}}
({K}^{ab} - {h}^{ab} {K} )
\end{equation}
where ${K}^{ab}$ is the extrinsic curvature and ${K}=
{K}^{ab}{h}_{ab}$.
The momentum conjugate to the scalar $\phi$ is
\begin{equation}\label{piphi}
\pi_{\phi} = \frac{\partial \mathcal{L}}{\partial
\dot{\phi}} = \beta \sqrt{{h}} \,
{n}^{\mu} \nabla _{\mu} \phi
\end{equation}
while the momentum conjugate to the $p$-form potential
$ {B}$ is
\begin{equation} \label{piB}
\pi_{B}^{a_{1} \ldots a_{p}} =
\frac{\partial
\mathcal{L}}{\partial
\dot{B}_{a_{1} \ldots a_{p}}} = \frac{\beta \sqrt{h}}{p!} e^{-\alpha
\phi} {n}_{\mu}
{H}^{\mu a_1 \ldots a_p}
\end{equation}

We define the Hamiltonian density canonically
 \be \label{Hdencan}
 \mathcal{H} = p^{i} \dot{q_{i}} - \mathcal{L}
 \ee
 where $\mathcal{L}$ is the Lagrangian density and $p^{i}$ and $q_{i}$ are respectively the time derivatives and momentums of the metric and fields, namely (\ref{piG})-(\ref{piB}).  In the above definition we discard any surface terms.   We will then add appropriate surface terms  to ensure the Hamiltonian has a well defined variational principle.  Alternatively, one could derive the same results by beginning with an action with the surface terms necessary to make its variation well defined and carrying these terms through the calculation.

 The volume Hamiltonian defined by integrating (\ref{Hdencan}) is
 \begin{equation} \label{Hconst}
{H}_{V}
 = \int_\Sigma \Big(\xi^{\mu}C_{\mu} + \xi^{\mu_{1}}B_{\mu_{1}
a_{2} \ldots
a_{p}}\mathcal{C}^{a_{2} \ldots a_{p}}\Big)
\end{equation}
where $C_{\mu}$ are the constraints from the Einstein equations and $
\mathcal{C}^{a_{2} \ldots a_{p}} $ the constraint from the
$p$-form.  For the sake of compactness we have adopted the convention that when $p=1$, $\mathcal{C}^{a_{2} \ldots a_{1}} = \mathcal{C}$ (i.e. a scalar).  Explicitly,
$$
C_{0} = -2\sqrt{h} (G_{\mu\nu} - 8\pi T_{\mu\nu})n^{\mu}n^{\nu}
= - \beta \sqrt{h} R^{({d} - 1)}
+ \frac{1}{ \beta \sqrt{h}}(\pi_{G}^{ab}\pi^{G}_{ab} +
\frac{\pi_{G}^2}{2 - {d}})
$$
$$
+ \frac{\pi_{\phi}^2}{2 \beta \sqrt{h}} + \frac{\beta \sqrt{h}}{2}(D
\phi)^2  + \beta \sqrt{h} V(\phi)
$$
\be
+ \frac{p!}{2 \beta \sqrt{h}} e^{\alpha \phi} \pi_{B}^2 + \frac{\beta \sqrt{h}}{2 (p+1)!} e^{-\alpha \phi} H_{a_{1} \ldots a_{p+1}}
H^{a_{1} \ldots a_{p+1}}
\ee
\begin{equation}
C_{a} =  -2\sqrt{h} (G_{a \mu} - 8\pi T_{a \mu}) n^{\mu}
=  -2\sqrt{h}D_{c} (
\frac{{\pi_{G}}^{c}_{a}}{\sqrt{h}}  ) +
\pi_{\phi}D_{a}\phi
+  \pi_{B}^{a_1 \ldots a_{p}} H_{a a_{1} \ldots a_{p}}
\end{equation}
\begin{equation} \label{gaugeconst}
\mathcal{C}^{a_{2} \ldots a_{p}} = -p\sqrt{h} D_{a}\Big(\frac{\pi_{B}^{a
a_{2} \ldots a_{p}}}{\sqrt{h}}\Big)
\end{equation}
In accordance with the above convention, in the case of $p=1$ (\ref{gaugeconst}) should be read as
\be
\mathcal{C} = -\sqrt{h} D_{a}\Big(\frac{\pi_{B}^{a}}{\sqrt{h}}\Big)
\ee
Note the ``pure constraint'' form of the volume Hamiltonian is no accident, but in fact is generic to any theory with time reparametrization invariance.  This follows from the fact that the Hamiltonian generates time translations.   Due to this form, $H_V$ will vanish if one considers any solution satisfying the constraints.

We have, however, yet to add the surface terms to obtan a well defined variational principle.  When one varies (\ref{Hconst}) and then performs integrations by parts to produce the equations of motion, a variety of surface terms are produced.  We must then add terms to cancel off these quantities.  Specifically one must add
$$
\beta \int dS^{a} \Big[ N D^b (\delta h_{a b}) - D^b(N) \delta h_{a b} + h^{c d}(-N D_a (\delta h_{c d}) + D_a (N) \delta h_{c d} \Big]
$$
$$ + \int dS_a \Big[ 2 N_b \frac{\delta {\pi_G}^{a b}}{\sqrt{h}}  + 2 N^c \frac{{\pi_G}^{a b}}{\sqrt{h}} \delta h_{b c} - N^a \frac{{\pi_G}^{b c}}{\sqrt{h}} \delta h_{b c} \Big] $$
$$+p \int dS_a \xi^{\mu} B_{\mu a_2 \ldots a_p} \delta \Big( \frac{{\pi_B}^{a a_2 \ldots a_p}}{\sqrt{h}} \Big)  - \int dS_a \Big( \beta N D^a \phi + N^a \frac{\pi_{\phi}}{\sqrt{h}} \Big) \delta \phi$$
\be \label{HS1}
 - \int dS_a \Big (\frac{\beta N}{p!} e^{-\alpha \phi} H^{a a_1 \ldots a_p} + \frac{p + 1}{\sqrt{h}} N^{[a} {\pi_B}^{a_1 \ldots a_p]} \Big) \delta B_{a_1 \ldots a_p}
 \ee
The pure matter terms have previously been written down in \cite{CopseyHorowitz} and the purely gravitational terms have been previously found by Regge and Teitelboim\cite{RT}.

Now we must find a series of finite surface terms to be added to the Hamiltonian such that the variation yields  (\ref{HS1}).   There does not seem to be a generic way to do this, but rather one must find the terms appropriate on a case by case basis for the desired asymptotics.  Since the solutions we are interested in are non-rotating, for the sake of simplicity we restrict ourselves to the case $N^a = 0$.  We wish to consider warped product solutions with $D = d + q$ dimensions
\be
ds^2 = k_{i j}(y) dy^i dy^j + l_{a b}(x,y) dx^a dx^b
\ee
where we impose the standard aymptotically AdS boundary conditions in the $y^i$ directions (see e.g. \cite{HIM})
\be \label{AdSBdy}
k_{i j} = k^{(0)}_{i j} + \delta k_{ i j} \Big(\frac{l}{r} \Big)^{d-1}
\ee
 with $k^{(0)}_{i j} $ the metric of global $AdS_{d}$
\be
k^{(0)}_{i j}  dy^{i} dy^{j} = -\Big(\frac{r^2}{l^2} + 1\Big) dt^2 + \frac{dr^2}{\frac{r^2}{l^2} + 1} + r^2 d \Omega_{d-2}
\ee
and
\be \label{SBdy}
l_{a b} = l^{(0)}_{a b} + \delta l_{ a b} \Big(\frac{l}{r} \Big)^{d-1}
\ee
where $l^{(0)}_{a b} $ is the metric of a compact manifold $\mathcal{M}_q$.  Of particular interest is the case where $\mathcal{M}$ is a q-sphere.  To ensure that the gravitational surface terms in the Hamiltonian are finite we require that  as a function of $r$
 \be \label{dl}
 \delta l_{ a b} = \mathcal{O}(1)
 \ee
 In accordance with the standard AdS boundary conditions we also require
$$
 \delta k_{t t} = \mathcal{O}(r^{2})
 $$
 $$
\, \, \,  \delta k_{t r} = \mathcal{O}(r^{-1})
 $$
 $$
\,  \delta k_{t \theta_i} = \mathcal{O}(r^{2})
 $$
 $$
\,  \, \, \delta k_{r r} = \mathcal{O}(r^{-2})
 $$
  $$
\, \, \, \,  \delta k_{r \theta_i} = \mathcal{O}(r^{-1})
 $$
 \be \label{dk}
\,  \, \,  \delta k_{\theta_i \theta_j} = \mathcal{O}(r^{2})
 \ee
 or, at least for the diagonal components, $\delta k_{i j}$ is of the same order as  $k^{(0)}_{i j} $.  We utilize the convention in (\ref{dl}, \ref{dk}) that the quantities are bounded above by the right hand side and may well be smaller or zero in particular circumstances.  These conditions then ensure that the gravitational surface terms in the Hamiltonian (\ref{HS1}) are finite and well defined.
 
 Note all of the subsequent analysis of this section will go through just as well if the q-dimensional manifold is absent.  In particular, if the dimensional reduction to d dimensions does not produce matter of types besides that in  (\ref{actn}) the results will apply equally well.  The analysis will also go through for orbifolded $AdS_p \times \mathcal{M}_q$ provided the structure we have assumed above is preserved and only the intervals of the various coordinates are affected.  The orbifold we will consider in the next section is precisely of this type.

The scalar term proportional to $N$ will be finite if asymptotically
\be
\phi - \phi_{0} \sim \frac{1}{r^{(d-1)/2}}
\ee
and vanish if $\phi$ falls off any faster.  Hence for scalars saturating the Breitenlohner-Freedman bound \cite{BF} which satisfy the fast (i.e. non-logarithmic) falloff rate this term will be nonzero and finite.  For other scalars the term will vanish if the fast fall-off rates are required.\footnote{As has been pointed out in recent years, one may study scalars just above the BF bound with slower falloff conditions provided one also weakens the above boundary conditions on the metric in such a way that divergences between this term and the gravitational terms cancel (\cite{HM}-\cite{AHHM}). Such boundary conditions appear to be perfectly sensible, although not the ones we wish to impose here.  }   In particular, in the case where one has a self-dual field $H$ (and hence $H^2 = 0$) and a vanishing scalar potential the field equation for $\phi$ is
\be
\nabla^2 \phi + \frac{\alpha e^{-\alpha \phi}}{2 (p+1)!} H^2 = \nabla^2 \phi = 0
\ee
and so, presuming $\phi$ approaches the constant $\phi_0$ asymptotically, by the usual analysis
\be
\phi - \phi_{0} \sim \frac{1}{r^{d-1}}
\ee
and the scalar will not contribute to the Hamiltonian.   Note it is also consistent in this case to set $\phi = 0$  and we will do so in the next section.

In terms of p-forms our primary interest is in the case where any electric field is of rank d and extended only in the y-directions (i.e. $H_{t y_2 \ldots y_d}$) and any magnetic field is of rank q and extended only in the x-directions.  In particular, the Freund-Rubin ansatz \cite{FR} we will be using later falls into this category.  Due to the fact that the magnetic field is closed, it is only a function of $x$,  $H_{x_1 \ldots x_q} (x)$, and hence the potential $B_{x_1 \ldots x_{q-1}}$ is independent of r and equal to its asymptotic value.  Since this is the same as the background value, there is no magnetic contribution to the energy in these cases.

With the above boundary conditions then the desired term and the on shell value of the Hamiltonian becomes
$$
H_S = \beta \int dS^{a} \Big[ N h^{b c} \Big( D_c (\Delta h_{a b}) -  D_a (\Delta h_{b c }) \Big) - D^b(N) \, \Delta h_{a b} + D_{a} (N)  \, h^{b c } \Delta h_{b c} \Big]
$$
\be \label{surfdone}
-  \frac{\beta}{2}   \int dS_aN  \phi  \, D^a \phi + (d - 1) \int dS_a \xi^{\mu} B_{\mu a_2 \ldots a_{d-1}} \Delta \Big( \frac{{\pi_B}^{a a_2 \ldots a_{d-1}}}{\sqrt{h}} \Big)
\ee
where $\Delta h_{a b} = h_{a b} - {h^{(0)}}_{a b}$ where $h^{(0)}$ is the background metric, $D_a$ is the covariant derivative with respect to the background metric, and indices are raised with the background metric.  Likewise, the remaining $\Delta$ indicates a subtraction from the background value of the indicated quantity.  The background subtractions imply that the energy of an undistorted $AdS_p \times \mathcal{M}_q$ vanishes with this definition.  Of course, any definition of energy will have a zero-point ambiguity  and this must be fixed either with comparison to a particular spacetime or renormalization prescription.  With the given normalization,  (\ref{surfdone}) is the unique set of surface terms that make the Hamiltonian well defined since the variation of the Hamiltonian is fixed.

We would now like to focus on the last electric term.
\be \label{Econt}
(d - 1) \int dS_a \xi^{\mu} B_{\mu a_2 \ldots a_{d-1}} \Delta \Big( \frac{{\pi_B}^{a a_2 \ldots a_{d-1}}}{\sqrt{h}} \Big)
\ee
Note this contribution does not look gauge independent.  In fact, it turns out there is only one physically acceptable choice of gauge in this circumstance.   We should emphasize here one would have such a term in the $AdS_5 \times S_5$ context regardless of whether one examines the situation from a five or ten dimensional perspective.  In particular the restriction we discuss below is not related to the well known fact that one may not write down a covariant lagrangian that ensures the self-duality of the field.  We take the usual solution to that problem and impose the self-duality by hand.  Alternatively, one might perform the dimensional reduction, but (\ref{Econt}) will remain as above.

The momentum must satisfy the constraint
\be \label{piconst}
D_{y_1}\Big(\frac{\pi_{B}^{y_1
\ldots y_{d-1}}}{\sqrt{h}}\Big) = \frac{1}{\sqrt{h}} \partial_{y_1} \Big(\pi_{B}^{y_1
\ldots y_{d-1}}\Big) = 0
\ee
Then, recalling that in this circumstance there is only one nonzero component of the momentum,
\be
\pi_{B}^{y_1 \ldots y_{d-1}}=\pi_{B}^{y_1 \ldots y_{d-1}} (t,x)
\ee
In fact $\pi$ is time independent as well, a fact ensured by the scalar constraint considering that the metric is asymptotically constant and the magnetic form $H$ is closed (and hence time independent).
Then
\be \label{momexp}
\frac{\pi_{B}^{y_1 \ldots y_{d-1}}}{\sqrt{h}} = \frac{C_1(x)}{\sqrt{h^{(0)}}} \Big(1 + \mathcal{O} \Big(\frac{l}{r} \Big)^{d-1} \Big)
\ee
where $h^{(0)}$ is the determinant of the spatial asymptotic metric (i.e. that of ${k^{(0)}_{i j }}$ at $t = 0$ and ${l^{(0)}}_{a b}$).
Then
\be
\Delta \Big( \frac{{\pi_B}^{y_1 y_2 \ldots y_{d-1}}}{\sqrt{h}} \Big) \sim r^{3-d} r^{1-d}
\ee
where the first factor comes from $\sqrt{{h^{(0)}}}$ and the second from the leading order corrections to the metric.
Since the integration measure grows as $r^{d-3}$, the term (\ref{Econt}) will yield a finite contribution if
\be \label{potcond}
\xi^{\mu}B_{\mu y_2 \ldots y_{d-1}} \sim r^{d-1}
\ee
and vanishes if $\xi \cdot B$ is any smaller.

The field strength corresponding to such momentum is
$$
H_{t r \theta_1 \ldots \theta_{d-2}} = C_2 \sqrt{-k^{(0)}}\Big(1 + \mathcal{O} \Big(\frac{l}{r} \Big)^{d-1} \Big)
$$
\be
= \partial_t (B_{r \theta_1 \ldots \theta_{d-2}}) - \partial_r (B_{t \theta_1 \ldots \theta_{d-2}}) + \Sigma_{\theta_i} \partial_{\theta_1} (B_{t r \theta_2 \ldots \theta_{d-2}})
\ee
where the last sum implicitly contains the appropriate signs for permutations and $C_2$ is a constant.  By far the most obvious choice of gauge is that only $B_{t \theta_1 \ldots \theta_{d-2}} \neq 0$ and hence
\be
B_{t \theta_1 \ldots \theta_{d-2}} = \frac{C_2}{1-d} \underbrace{r\sqrt{-k^{(0)}}}_{\sim r^{d-1}} \Big(1 + \mathcal{O} \Big(\frac{l}{r} \Big)^{d-1} \Big)
\ee
and so (\ref{Econt}) is finite.  Note that in this context $\Delta (\pi/\sqrt{h})$ is not the variation of the charge; the electric flux is fixed asymptotically while $\Delta (\pi/\sqrt{h})$  describes the leading order corrections to the flux.  In fact, one might well worry it is not even gauge invariant, as it is inversely proportional to the square root of a determinant.  One can show this concern is well justified as follows. By making the coordinate reparametrization
\be \label{reparam}
r = \bar{r}\Big(1+ a \Big(\frac{l}{\bar{r}}\Big)^{d-1}\Big)
\ee
one changes the value of the coefficients of the leading order corrections to the metric but leaves the form intact.  The changes due to this reparametrization in terms subleading in (\ref{momexp}) will not be large enough to change the surface term (\ref{Econt}), but $h^{(0)}$ will be altered.  Specifically one finds
\be
\sqrt{h^{(0)}} = \sqrt{\bar{h}^{(0)}} \Big(1- a \Big(\frac{l}{\bar{r}}\Big)^{d-1} + \mathcal{O}(\bar{r}^{2 - 2d}, \bar{r}^{-d-1})\Big)
\ee
where $\bar{h}^{(0)}$ is the determinant of the asymptotic spatial metric in terms of $\bar{r}$.
Then $\Delta (\pi/\sqrt{h})$ is not invariant under this reparametrization and neither is  the surface term (\ref{Econt}), since the remainder of the terms only are relevant at leading order.   On the other hand, it is straightforward to show the gravitational terms are invariant under this change.   The latter should not be surprising; the  calculation is identical to the one one would do to verify that the mass of a solution which is asymptotically $AdS_d$ (instead of $AdS_d \times \mathcal{M}_q$) is invariant under this change of coordinates.  While it seems likely there is a generic underlying explanation for this pathology, for the present we simply note its existence and consider other possible gauge choices.

Since the field strength is proportional to the volume form on the sphere, if one made a choice of gauge such that $B_{t r \theta_1 \ldots \theta_{d-3}} \neq 0$ one necessarily would have to define the  potential in patches.   This follows simply because the volume form on the sphere is not exact.  While such a gauge choice can eliminate the surface term at infinity, it does so at the cost of introducing integrals along the interface of the patches.  For the sake of illustration, call the potential in one patch ${B^{(1)}}_{t r \theta_1 \ldots \theta_{d-3}}$ and in a second (there may be several such patches) ${B^{(2)}}_{t r \theta_1 \ldots \theta_{d-3}}$.   Each patch has its own set of surface terms to make the Hamiltonian well defined in that patch.  Note that two patches which touch have opposite pointing normals ($\hat{n}^{(1)}$ and $\hat{n}^{(2)}$, respectively) and  so combining these terms one produces an integral over the interface of the two patches and the difference in the gauges:
$$
(d-1) \int_{1} {\hat{n}^{(1)}}_a  \, \xi^{\mu} {B^{(1)}}_{\mu a_2 \ldots a_{d-1}} \Delta \Big( \frac{{\pi_B}^{a a_2 \ldots a_{d-1}}}{\sqrt{h}} \Big)
$$
$$
 +  \, (d-1) \int_{2} {\hat{n}^{(2)}}_a  \xi^{\mu} {B^{(2)}}_{\mu a_2 \ldots a_{d-1}} \Delta \Big( \frac{{\pi_B}^{a a_2 \ldots a_{d-1}}}{\sqrt{h}} \Big)
$$
\be \label{intint}
= (d-1) \int_{1} {\hat{n}^{(1)}}_a  \, \xi^{\mu} \, ({B^{(1)}}_{\mu a_2 \ldots a_{d-1}} - {B^{(2)}}_{\mu a_2 \ldots a_{d-1}} ) \, \Delta \Big( \frac{{\pi_B}^{a a_2 \ldots a_{d-1}}}{\sqrt{h}} \Big)
\ee
 In addition to being inconvenient, this choice of gauge does not yield a finite Hamiltonian;
\be
B_{t r \theta_1 \ldots \theta_{d-2}} \sim r^{d-3}
\ee
and the jump in gauge is of the same order.  Then, since  (\ref{intint})  includes an integral over $r$ it is logarithmically divergent.

The only remaining possible gauge choice is the time dependent gauge
\be
B_{r \theta_1 \ldots \theta_{d-2}}  = C_2 (t - t_0) \sqrt{-k^{(0)}}\Big(1 + \mathcal{O} \Big(\frac{l}{r} \Big)^{d-1} \Big)
\ee
where $t_0$ is an arbitrary but fixed constant.  Note this makes the electric term in the Hamiltonian vanish (\ref{Econt}) and we are left with only the gravitational and dilatonic terms.  More geometrically, one may demand that asymptotically
\be\label{gauge1}
\xi \cdot B = 0
\ee
or
\be \label{gauge2}
n \cdot B = 0
\ee
We have listed the right hand sides of (\ref{gauge1}) and (\ref{gauge2}) as zero, although we will not run into any difficulty in the above case as long as they are not as large as $r^{d -1}$.  While we are not aware of any problems caused by enforcing the stronger conditions, neither do we have a generic argument that such difficulties can never occur.  As we discuss below, in other cases there is good reason to enforce (\ref{gauge1}) or (\ref{gauge2}) as stated.  In the case of electric fields of rank $d$ and a shift with no component along the compact (e.g. $S_5$) manifold these two conditions (\ref{gauge1})  and  (\ref{gauge2})  are equivalent;  any contribution due to the shift will vanish since all the angles in the space already appear contracted with $\pi$.  As we discuss later, however, one might hope to resolve this ambiguity by considering a more generic situation.  While it seems somewhat odd that even in a time independent background one is forced to choose a time dependent gauge, as noted above all other choices lead to an ill-defined Hamiltonian.   It would be interesting to understand what, if any, restriction this corresponds to in the gauge theory of the AdS-CFT correspondence.

For electric fields of lower rank (\ref{Econt}) will not have the same difficulties as above but it still may be finite.  In particular, consider asymptotically $AdS_d$ spacetimes with only a simple radial electric field $\pi^r(r)$.   Then
\be
\partial_r \pi^r = 0
\ee
and the solution will have an electric charge
\be
Q =\frac{1}{\Omega_{d-2}} \int dS_a \frac{\pi^a}{\sqrt{h}}
\ee
Then
\be
\delta \Big(\frac{\pi^r}{\sqrt{h}} \Big)  \sim \frac{\delta Q}{\sqrt{h}} \sim \frac{\delta Q}{r^{d-3}}
\ee
and
\be
H_{r t} \sim \frac{Q}{r^{d-2}}
\ee
The gravitational surface terms in this case will be just as they were above (\ref{surfdone}) but the electric term is now
\be \label{Econt2}
\int dS_a \, \xi^{\mu} B_{\mu} \, \frac{\pi^a}{\sqrt{h}}
\ee

If one takes a time independent gauge
\be
B_{t} = \varphi_0+ \mathcal{O}(r^{3-d})
\ee
unless the value of the potential at infinity ($\varphi_0$) is set to zero, (\ref{Econt2}) will be finite and nonzero.   This in fact should not be any surprise; this term yields the electric work term $\Phi \delta Q$ in the context of the first law of black hole thermodynamics \cite{CopseyHorowitz, Wald}.  Note changing the value of $\varphi_0$ is not just gauge but in fact would require doing work on the system.   In fact, if one is allowed to change $\varphi_0$ at will the value of the Hamiltonian for any charged system may be set to any desired value, positive or negative.  

By using the BPS bound for a rotating supersymmetric solution with an electric field with rank less than d one should be able to determine whether (\ref{gauge1}) or (\ref{gauge2}) is correct.  Unfortunately, the only suitable solutions we are aware of are rather complicated supersymmetric black holes (see \cite{HarveyRoberto} for a recent review)  where one needs to take into account not only surface terms at infinity but also a substantial number of surface terms at the horizon.  This appears to be technically rather involved and we will postpone it for future work.

Before finishing, we should note it has been asserted in the literature \cite{HIM} that under ``natural'' boundary conditions a p-form field will not contribute to the Hamiltonian for spaces that are asymptotically $AdS_d$ without any restriction on the gauge.  However, one can check that the boundary conditions imposed there imply a faster falloff than is physically required and in particular exclude any solutions with net global electric charge.  Hence the relevant terms in the Hamiltonian are ruled out by hand.  Of course, once one imposes the described gauge boundary conditions, the electric terms in the Hamiltonian (\ref{Econt}) and (\ref{Econt2}) will vanish and we agree with the final results of \cite{HIM}.

 \setcounter{equation}{0}
\section{Initial Data for $AdS_5 \times S^{5}/{Z_k}$}

Let us now turn to orbifolded $AdS_5 \times S^{5}$.  In particular we would like to consider the nonsupersymmetric orbifold with no fixed points described by \cite{HOP}.   In this example, the orbifold acts by equal rotations of $2\pi/k$ in each of the three orthogonal planes of the $S_5$.  This may be implemented by considering the $S_5$ as a Hopf fibration of $S_1$ over $CP_2$; the orbifold acts to reduce the length of the $S_1$ cycle by a factor of $k$.   For reasons described in detail in \cite{HOP}, there will be no tachyons provided $g_s N \gg k^4$ and $k$ is odd.  For $k \geq 5$ the authors of \cite{HOP} describe an instability while for $k=3$ the orbifold turns out to be supersymmetric and no such instability is found.

We wish to search for a far worse instability than that of \cite{HOP}--namely the existence of negative energy states.  One expects the lowest energy configurations at a given moment will be time symmetric initial data since in that case no energy is present in the form of gravitational momentum.  Further, a cross section of an instanton describing any decay of the vacuum must correspond to such data.   Thus we search here for suitable time symmetric initial data.  We may parametrize the $S_5$ by the three complex coordinates $(z^1, z^2, z^3)$ which satisfy $z^i \bar{z}^i = 1$ and $d {\Omega_5}^2 = dz^i d \bar{z}^i$.  These $z^i$ may be written as
$$
z^1 = e^{i (\phi_1 + \chi)} \cos \theta
$$
$$
\, \, \, \, \, \, \, \, \, \, \, \, \, z^2 = e^{i (\phi_2 + \chi)} \sin \theta \, \cos \psi
$$
\be \label{zparam}
\, \, z^3 = e^{i \chi} \sin \theta \, \sin \psi
\ee
The metric on ${CP}_2$ may be written in terms of four one forms 
\be
{ds^2}_{CP_2} = \Sigma_{a = 1}^{4} e_a e_a
\ee
and then 
\be \label{S5param}
d {\Omega_5}^2 = {ds^2}_{{CP}_2} + {e_5}^2
\ee
In terms of the coordinates of (\ref{zparam}), the one forms $e_a$ are
$$
 \, \, \,  \, \, \, e_1 = d\theta
$$
$$
 \, \, \,  \, \, \, \, \, \, \, \, \, \, \, \, \, \, \, \,e_2 = \sin \theta \, d\psi
$$
$$
 \, \, \,  \, \, \, \, \, \, \, \, \, \, \, \, \, \, \, \, \,\, \, \, \, \, \, \, \, \, \, \, \, \,\, \, \, \, \, \, \, \, \, \, \, \, \,\, \, \, \, \, \, \, \, \, \, \, \, \,\, \, \, \, \, \, \, \, \, \, \, \, \,e_3 = \sin \theta \, \cos \theta \, (d\phi_1 - \cos^2 \psi \, d \phi_2)
$$
\be
 \, \, \,   \, \, \, \, \, \, \, \, \, \, \, \, \, \, \, \, \, \, \, \, \,\, \, \, \, \, \, \, \, \, \, \, \, \,\, \, \, \, \, \, \, \, \, \, \, \, \,e_4 = \sin \theta \, \sin \psi \, \cos \psi \, d\phi_2
\ee
$$
\,\, \, \, \, \, \, \, \, \,\, \, \, \, \, \, \, \, \, \, \, \, \,\, \, \, \, \, \, \, \, \, \, \, \, \,\, \, \, \, \, \, \, \, \, \, \, \, \,\, \, \, \, \, \, \, \, \, \, \, \, \,\, \, \, \, \, \, \, \, \, \, \, \, \,\, \, \, \, \, \, \, \, \, \, \,  
e_5 = d\chi + \cos^2 \theta \, d \phi_1 + \sin^2 \theta \cos^2 \psi \, d \phi_2
$$
For the original $S_5$ the period of $\chi$ is $2 \pi$.  In the $Z_k$ orbifold the form of (\ref{S5param}) is not altered but $\chi$ has period $2 \pi/k$.

We wish to consider bubbles produced when the $\chi$ cycle pinches off.  The simplest such initial data is
\be \label{ansatz}
ds^2 =  \frac{dr^2}{W(r)} + f(r) d\Omega_3 + g(r) \, {ds^2}_{CP_2}  + h(r) \, {e_5}^2
\ee
Of course, one of these functions of $r$ is pure gauge, but it turns out to be convenient to leave the gauge unfixed for the present.   In terms of the coordinates of (\ref{zparam}), (\ref{ansatz}) is
$$
ds^2 = \frac{dr^2}{W(r)} + f(r) d\Omega_3 + g(r) \Big(d \theta^2 + \sin^2 \theta d \psi^2 \Big)
$$
$$
+ h(r) \Big( d\chi^2 + 2 \cos^2 \theta \, d \chi d \phi_1 + 2 \sin^2 \theta \cos^2 \psi \, d \chi d \phi_2 \Big)
$$
$$
+ \cos^2 \theta \Big( \cos^2 \theta \, h(r) + \sin^2 \theta \, g(r) \Big) d\phi_1^2
$$
$$
+ 2 \sin^2 \theta \cos^2 \theta \cos^2 \psi \Big( h(r) - g(r) \Big) d\phi_1 d \phi_2
$$
\be
+ \sin^2 \theta \cos^2 \psi \Big( h(r) \sin^2 \theta \cos^2 \psi + g(r) (\cos^2 \theta \cos^2 \psi + \sin^2 \psi) \Big) d\phi_2^2
\ee
In particular a $t = 0$ slice of the solutions of \cite{HOP} falls into this class.

The only matter for the solutions we wish to consider is a self-dual five form
\be
F_5 = \xi(r) ( \epsilon_5 + \star \epsilon_5)
\ee
where $\epsilon_5$ is the volume form on the $S_5$.  The requirement that the magnetic field  (or equivalently $F $) is a closed form implies that
\be \label{Fr}
\xi(r) = \frac{C_3}{g^2(r) \sqrt{h(r)} }
\ee
for a constant $C_3$.  Matching the value of $C_3$ to its asymptotic value determines
\be
C_3 =  \sqrt{8} R^4
\ee
where $R$ is the asymptotic radius of the $S_5$ (i.e. $g(\infty) = h(\infty) = R^2$).  Given (\ref{Fr}) the gauge constraint is satisfied and one only needs consider the scalar constraint
\be \label{const}
{}^{(9)} R =  \frac{1}{2} \Big( \frac{E^2}{4!} + \frac{\bar{F}^2}{5!} \Big)
\ee
where ${}^{(9)} R$ is the scalar curvature constructed from the initial data, $\bar{F}$ is the form field projected into the initial data surface and
\be
E_{a_1 a_2 a_3 a_4} = n^{\mu} F_{\mu a_1 a_2 a_3 a_4}
\ee
with $n^{\mu}$ is a unit timelike vector orthogonal to the initial data surface.  Then the scalar constraint becomes
\be \label{constexp}
{}^{(9)} R = \frac{C_3^2}{g^4(r) h(r)}
\ee
Inserting the given form of the metric then the constraint (\ref{const}) yields
\be \label{const2}
a_0(r) W'(r) + a_1(r) W(r) = a_2(r)
\ee
where
\be
a_0(r) =\frac{3 f'(r)}{2 f(r)} + \frac{2 g'(r)}{g(r)} + \frac{h'(r)}{2 h(r)}
\ee
$$
a_1(r) =  \frac{6 f'(r) g'(r)}{f(r) g(r)} + \frac{{g'(r)}^2}{g(r)^2} + \frac{3 f'(r) h'(r)}{2 f(r) h(r)} + \frac{2 g'(r) h'(r)}{g(r) h(r)} - \frac{{h'(r)}^2}{2 h(r)^2}
 $$
$$
  + \frac{3 f''(r)}{f(r)} +  \frac{4 g''(r)}{ g(r)}  +  \frac{h''(r)}{ h(r)}
$$
\be
= 2 a_0'(r) + a_0 (r) \frac{h'(r)}{h(r)} + a_3 (r)
\ee
where
\be
a_3(r) = 3\Big(\frac{g'(r)}{g(r)} + \frac{f'(r)}{f(r)} \Big)^2 + 2 \Big(\frac{g'(r)}{g(r)} \Big)^2
\ee
and
\be
a_2(r) = \frac{6}{f(r)} + \frac{24}{g(r)}  - \frac{4 h(r)}{g^2(r)} - \frac{C_3^2}{g^4(r) h(r)}
\ee
One may then choose $f(r)$, $g(r)$, and $h(r)$ arbitrarily and the constraint (\ref{const2})  is solved by taking
\be \label{sol1}
W(r) = \frac{e^{-\int_{r_0}^{r} du \frac{a_3 (u)}{a_0(u)}}}{a_0^2(r) h(r)} \Bigg[ C_4+ \int_{r_0}^{r} ds \, a_0 (s) h(s) a_2 (s) \, e^{\int_{r_0}^{s} dt \frac{a_3 (t)}{a_0(t)}} \Bigg]
\ee
The constant $C_4$ will be used below to ensure the absence of a conical singularity.

We would like to consider bubble solutions of size $r_0$ (i.e. $h(r_0) = 0$).  There are no entirely regular solutions but there are solutions, like those in \cite{HOP}, where the metric approaches that of a stack of D3 branes wrapped around the $S_3$ of the $AdS_5$ space and smeared over the $CP_2$.\footnote{An unsmeared stack of D3 branes would not be singular but the smeared stack is.  This may be seen by noting via (\ref{constexp}) the fact that the scalar curvature diverges at the surface of the bubble (where $h(r)$ vanishes).}  Since this leaves only two directions orthogonal to the branes, the appropriate harmonic function is a logarithm.  Then if one defines
\be
\gamma(r) = \sqrt{-\log\Big(\frac{r}{r_0} - 1 \Big)}
\ee
 we seek a solution such that for $r \sim r_0$
\be \label{nearbrane}
f \sim \frac{f_0}{\gamma}, \, \, \, \, \, g \sim g_0 \gamma, \, \, \, \, \, \, h \sim h_0 (r - r_0) \gamma
\ee
and
\be
W \sim w_0 \frac{(r - r_0)}{\gamma}
\ee
for some constants $f_0, g_0, h_0$ and $w_0$.  Once we demand the above behavior for the functions $f$, $g$, and $h$ the form of $w$ follows from the constraint (\ref{sol1}).
To see this, if one takes the prescribed form for $f$, $g$, and $h$, for $r \sim r_0$
\be
\frac{a_3}{a_0} \sim \frac{1}{\gamma^4 (r - r_0)}
\ee
and so
\be
\int_{r_0}^{s} du \frac{a_3 (u)}{a_0(u)} \sim \frac{1}{\gamma^2(s)}
\ee
Further
\be
a_0 a_2 h \sim - \frac{4 R^8}{g_0^4} \frac{1}{\gamma^4 (r - r_0)}
\ee
and thus
\be
 \int_{r_0}^{r} ds \, a_0 (s) h(s) a_2 (s) \, e^{\int_{r_0}^{s} du \frac{a_3 (u)}{a_0(u)}}  \sim -\frac{4 R^8}{g_0^4} \frac{1}{\gamma^2(r)}
\ee
Hence we are justified in taking $r_0$ in (\ref{sol1}) as the bubble size (in contrast to the situation if, for example, the integral of $ a_3/a_0$ or of $ a_0 a_2 h$ diverged as  $r \sim r_0$).  Then one finds W has the desired form with
\be
w_0 = \frac{4 C_4}{h_0}
\ee
There will be no conical singularity provided one takes
\be
C_4= k^2
\ee

We are interested in solutions which are asymptotically $AdS_5 \times S_5/Z_k$.  It is straightforward using (\ref{sol1}) to check that provided that $f(r)$ satisfies the usual asymptotically AdS requirement, namely
\be
f(r) = r^2 \Big(1 + \delta f  \frac{l^4}{r^4} + \mathcal{O}(r^{-4 - \epsilon}) \Big)
\ee
for constant $\delta f$ and positive $\epsilon$, and $g(r)$ and $h(r)$ fall off quickly enough to make the Hamiltonian well defined
\be
g(r) = R^2 \Big(1 + \delta g \frac{l^4}{r^4} + \mathcal{O}(r^{-4 - \epsilon}) \Big)
\ee
and
\be
h(r) = R^2 \Big(1 + \delta h \frac{l^4}{r^4} + \mathcal{O}(r^{-4 - \epsilon}) \Big)
\ee
where $\delta g$ and $\delta h$ are likewise suitable constants
that $W(r)$ will satisfy the usual asymptotically AdS requirement, namely
\be
W(r) = \frac{r^2}{l^2} + 1 - \delta_{r r} \frac{l^2}{r^2} + \mathcal{O}(r^{-2 - \epsilon})
\ee
for some constant $\delta _{r r}$.  Using the definition of mass  developed in the previous section (and taking the conventional normalization $\beta = 1/{16 \pi G_D}$)  these solutions will have energy
\be
M = \frac{R^7}{16 \pi G_{10}} (3 \, \delta_{r r} + 12 \,  \delta f + 20 \delta g + 5 \,  \delta h)\times \Omega_3 \times \frac{\Omega_5}{k}
\ee

 \setcounter{equation}{0}
\section{Negative Energy Solutions}

We now wish to see if there are any negative energy solutions of the constraint.  Unfortunately, choosing any $f, g$, and $h$ such that $W$ may be written explicitly seems quite difficult  for functions having the required behavior near the bubble (\ref{nearbrane}).  However, one may find simple $f, g$, and $h$  such that the relevant integrals for $W$ may be found in certain regions.  We  may then patch together such solutions in order to find a sufficently explicit form of $W$.  Note that in terms of finding the asymptotic value of $W$, it will only matter if the resultant solution is $C_0$; the integrals in (\ref{sol1}) only involve first derivatives and so any smoothness beyond continuity will not make any contribution.   To be precise, if one introduces smoothing over some small region $\delta$, that smoothing will make an $\mathcal{O}(\delta)$ contribution to the energy.  However, to remove any doubt from the most careful reader's mind we will only consider initial data that is $C_2$; one may, as noted above, smooth this to any desired degree at an arbitrarily small cost in mass.  Finally, note via (\ref{sol1}) if $f, g$, and $h$  are $C_N$ then $W$ is $C_{N-1}$.

The simplest possible case is that in which the functions in the brane region match smoothly onto the asymptotic values, i.e.
\be \label{asym1}
g(r) = h(r) = R^2
\ee
and
\be \label{asym2}
f(r) = r^2
\ee
where $R$ is the asymptotic value of the radius of the $S^5$.  Hence consider for $r_0 \leq r \leq r_1$
\be
f = \frac{r_1^2 \gamma(r_1)}{\gamma(r)} , \, \, \, \, \, \, g = \frac{R^2 \gamma(r)}{\gamma(r_1)}, \, \, \, \, \, h = \frac{R^2 \gamma(r)}{\gamma(r_1)} \frac{r - r_0}{r_1 - r_0}
\ee
and for $r \geq r_2 \, $ $f, g$, and $h$  take their asymptotic values (i.e. (\ref{asym1})-(\ref{asym2})).  One may then patch these functions together in a $C_3$ fashion, that is for $r_1 < r < r_2$ (where $r_2/r_1 - 1 \ll 1$)
\be \label{match1}
f(r) = r^2 (1 + \Sigma_{i = 4}^{7} f_i (r - r_2)^i) , \, \, \, \, \, \, g(r) = R^2(1 + \Sigma_{i = 4}^{7} g_i (r - r_2)^i)
\ee
and
\be \label{match2}
h(r) = R^2(1 + \Sigma_{i = 4}^{7} h_i (r - r_2)^i)
\ee
where the constants $(f_i, g_i, h_i)$ are determined by the matching conditions.  It is straightforward to show that
\be
f_i = \mathcal{O}( (r_2 - r_1)^{-i + 1})
\ee
and analogous statements for $g_i$ and $h_i$ hold, so for $r_1 < r < r_2$, $f \approx r^2$, $g \approx R^2$, and $h \approx R^2$ while the first derivatives are of order $r_1$ and $R^2/r_1$ respectively.   Then, provided no derivatives become large or $a_0$ becomes small,  the integrals in (\ref{sol1}) in the matching region make only an $\mathcal{O}(r_2/r_1 - 1)$ contribution to the mass.  This contribution may then be made as small as desired.   We will show below that the above mentioned restrictions turn out to be easy to satisfy.  As $r_2$ approaches $r_1$ one will find curvatures of order $r_1^{-1} (r_2 - r_1)^{-1}$, but as long as $r_1$ and $r_2$ are large compared to the Planck scale the classical analysis will still be reliable.  It turns out, as shown in detail below, that the solutions with large negative mass occur when $r_1$ becomes large, so for the most relevant and dangerous states the curvature in the matching region becomes parametrically small compared to the Planck scale.

Note that $r_1/r_0$ may not be taken to be arbitrarily large, since the reality of $\gamma(r)$  implies $r_1/r_0 < 2$.  In fact, the requirement that we have a regular solution imposes a stronger constraint.   For $r_0 < r <  r_1$
\be
a_0(r) = \frac{1}{2 ( r - r_0)} \Bigg(1 + \frac{1}{\log\Big(\frac{r}{r_0} - 1 \Big)} \Bigg)
\ee
and one must require this remains positive (otherwise $W(r)$ would diverge, at least generically, when $a_0$ vanishes -- see  (\ref{sol1})) and hence that
\be \label{rest1}
1 < \frac{r_1}{r_0} < 1 + \frac{1}{e} \approx  1.368
\ee
As it turns out, the interesting case for this class of examples falls well within this restriction.

For $r_0 < r < r_1$ one finds
$$
W(r) = \frac{4 (r_1 - r_0) \gamma(r_1)}{R^2} \frac{(r - r_0) \, \gamma(r)}{\Big(-\log\Big(\frac{r}{r_0} - 1 \Big)-1\Big)}
$$
$$
\Bigg[ k^2 +\frac{4 \log^2 \Big(\frac{r_1}{r_0} - 1 \Big)}{\log{ \Big(\frac{r}{r_0} - 1 \Big)}}+  \frac{(r - r_0)}{(r_1 - r_0)} \Big(12 -  \frac{(r - r_0)}{(r_1 - r_0)} \Big)
$$
\be
+ \frac{3 R^2}{{r_1}^2 \gamma^2(r_1)}  \frac{(r - r_0)}{(r_1 - r_0)}\Big(1-\log \Big(\frac{r}{r_0} - 1 \Big) \Big) \Bigg]
\ee
Then provided (\ref{rest1}) is enforced $W(r)$ will be regular in this range.  We also must ensure that $W(r)$ never goes through a zero in this region.  For small bubbles this is manifest while for large bubbles there is an additional restriction.  One simple sufficient, but not necessary, criterion is that
\be
k^2 + 4 \log \Big( \frac{r_1}{r_0} - 1 \Big) > 0
\ee
Combining this with (\ref{rest1}) implies
\be  \label{rest2}
1+e^{-\frac{k^2}{4}} < \frac{r_1}{r_0} < 1 + \frac{1}{e}
\ee
It turns out this is sufficient for our purposes.

For the matching region the form of $W(r)$ apparently cannot be obtained explicitly.   In the limit that $r_2 \rightarrow r_1$, provided $a_0$ does not become very small and $R/r_1$ is bounded, the integral contributions to W will make a small change to its value at $r = r_1$.   In this regime $h$ is approximately constant, so the only factor that can significantly change $W$ is $a_0$.   Note that $a_0$ is not necessarily approximately constant in the matching region.  There are even values of the parameters where it goes through a zero.   The minimum restriction that we must make is that $a_0$, as computed from (\ref{match1}) and (\ref{match2}), is positive definite when $r_2/r_1 - 1$ becomes arbitrarily small.  This is equivalent to the statement that
\be \label{lim2}
1.047 \lessapprox \frac{r_1}{r_0} \lessapprox 1.367
\ee
independent of k.  Under the modest additional restriction
\be \label{lim3}
1.050 \lessapprox \frac{r_1}{r_0} \lessapprox 1.303
\ee
one finds $a_0$ is of order one, precisely
\be
\frac{1}{4} \lessapprox a_0 \lessapprox 7
\ee
for
\be 1 \lessapprox \frac{r_2}{r_1}  \lessapprox 1.1
\ee
 If $k > 3.50$ then both (\ref{lim2}) and (\ref{lim3}) are stronger than (\ref{rest2}).  It is worth noting that the restrictions  (\ref{lim2}) and (\ref{lim3}) are a result of the simple matching functions we have chosen ((\ref{match1}) and (\ref{match2})) rather than anything fundamental.  At the cost of more complicated matching functions the restrictions of (\ref{lim2}) and (\ref{lim3}) could be eliminated, but we will be able to find plenty of solutions within their bounds.  Henceforth, we will impose (\ref{lim3}) and  (\ref{rest2}), as well as assuming $r_2/r_1 -1 \ll r_1/R$.

For $r > r_2$ one finds
\begin{eqnarray}
W(r) &=& \frac{r_2^4}{9 R^2 r^2} \frac{1 + \log(\frac{r_1}{r_0} - 1)}{\log(\frac{r_1}{r_0} - 1)} \underbrace{e^{-\int_{r_1}^{r_2} du \, \frac{a_3(u)}{a_0(u)}}}_{I_0}\nonumber
\\
&&\Bigg[ k^2 +11+ 4  \log(\frac{r_1}{r_0} - 1) + \frac{3 R^2}{r_1^2} \Big(1 - \frac{1}{ \log(\frac{r_1}{r_0} - 1)} \Big) \nonumber
\\
&\quad&+  \frac{\log(\frac{r_1}{r_0} - 1)} {1 + \log(\frac{r_1}{r_0} - 1)} \Big[ \underbrace{\int_{r_1}^{r_2} ds \, a_0(s) a_2(s) h(s) e^{\int_{r_1}^{s} dt \, \frac{a_3 (t)}{a_0 (t)}}}_{I_1} \nonumber
\\
&\quad&
+ 9 \, \frac{R^2}{r_2^4} e^{\int_{r_1}^{r_2} dt \, \frac{a_3 (t)}{a_0 (t)}} \Big(r^2 \Big(1+\frac{r^2}{R^2} \Big) - r_2^2 \Big(1+\frac{r_2^2}{R^2} \Big) \Big) \Big] \Bigg] \nonumber
\\
&=& \frac{r^2}{R^2} + 1 - \frac{r_1^2}{3 r^2} \Big(2 + \frac{1}{\log^2(\frac{r_1}{r_0} - 1)}  + 3 \frac{r_1^2}{R^2}+ \mathcal{O} \Big(\frac{r_2}{r_1} - 1\Big) \Big)
\\
&\quad&
+ \frac{r_1^4}{9 R^2 r^2} \Big(1 + \frac{1}{ \log(\frac{r_1}{r_0} - 1)}\Big) \Big[ k^2 + 11 + 4 \log(\frac{r_1}{r_0} - 1) + \mathcal{O}\Big(\frac{r_2}{r_1} - 1\Big) \Big] \nonumber
\end{eqnarray}
where we have noted that both $I_0$ and $I_1$ are of order $r_2/r_1 - 1$ since they involve integrals of bounded functions over a vanishingly small range.   For small bubbles ($r_1/R \ll 1$) the mass is given by
\be
M = \frac{\Omega_3 \Omega_5 R^7 {r_1}^2 }{16 \pi k \, G_{10}} \Bigg(2 + \frac{1}{\log^2(\frac{r_1}{r_0} - 1)} \Bigg) \Bigg(1 + \mathcal{O}\Big(\frac{r_2}{r_1} - 1, \frac{r_1^2}{R^2} \Big) \Bigg)
\ee
and so is positive definite, as one might well expect.  On the other hand for large bubbles ($r_1/R \gg 1$)
\be
M = \frac{\Omega_3 \Omega_5 R^5 r_1^4 }{16 \pi k \, G_{10}} M_0 \Bigg(1 + \mathcal{O}\Big(\frac{r_2}{r_1} - 1, \frac{R^2}{r_1^2} \Big) \Bigg)
\ee
where
\be
M_0 = 3 - \frac{1}{3} \Big(1 + \frac{1}{ \log(\frac{r_1}{r_0} - 1)}\Big) \Big[ k^2 + 11 + 4 \log(\frac{r_1}{r_0} - 1) \Big]
\ee
$M_0$ is positive if $k \leq 3$, but if $k \geq 4$ there is a region where it becomes negative. To be precise, $M_0$ is negative if
\be \label{Mlims}
y_1 < \frac{r_1}{r_0} < y_2
\ee
where
\be
y_1 =  1 + e^{\frac{1}{8} \Big( -k^2 - 6 - \sqrt{k^4 - 4 k^2 - 140} \Big)}
\ee
and
\be
y_2 =  1 + e^{\frac{1}{8} \Big( -k^2 - 6 + \sqrt{k^4 - 4 k^2 - 140} \Big)}
\ee
In particular for $k = 4$, if
\be
1.026 \lessapprox \frac{r_1}{r_0}  \lessapprox  1.157
\ee
then $M_0$ will be negative, as shown in figure 1.
\begin{figure}
\begin{picture} (0,0)
    	\put(-110,5){$M_0$}
         \put(130, -104){$\frac{r_1}{r_0}$}
    \end{picture}
    \centering

	\includegraphics[scale= 1]{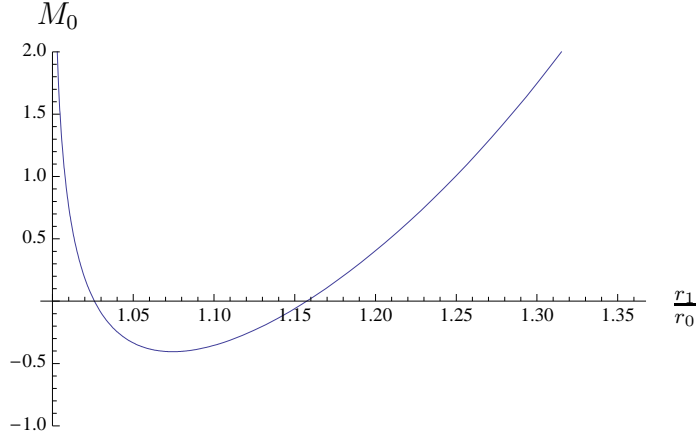}
	\caption{$M_0$ for $k = 4$}
	\label{ex1fig}
	\end{figure}

$M_0$ is qualitatively the same for any larger $k$.  As $k$ becomes larger, $y_1$ becomes smaller and approaches 1.  For $k \gg 1$ one finds
\be
y_1 \approx 1 + e^{-\frac{k^2}{4}}
\ee
Meanwhile, as $k$ increases, so does $y_2$.  For $k \gg 1$,
\be
y_2 \approx 1 + e^{-1 - \frac{9}{k^2}}
\ee
The minimum value of $M_0$ occurs at
\be \label{M0min}
\frac{r_1}{r_0} = 1 + e^{-\frac{\sqrt{11 + k^2}}{2}}
\ee
with value
\be
{M_0}_{\mathrm{min}} = \frac{-k^2 + 4 \sqrt{k^2+11} - 6}{3}
\ee
It is also worth noting if we choose any fixed $r_1/r_0$ such that $y_1 < r_1/r_0 < y_2$ at large $k$ the mass for this family of solutions becomes large and negative
\be
M = -\frac{k  r_1^4}{3} \Big(1 + \frac{1}{ \log(\frac{r_1}{r_0} - 1)}\Big) \frac{\Omega_3 \Omega_5 R^5}{16 \pi G_{10}} M_0 \Bigg(1 + \mathcal{O}\Big(\frac{r_2}{r_1} - 1, \frac{R^2}{r_1^2}, k^{-2} \Big) \Bigg)
\ee

Some of the above solutions correspond to solutions which are singular in the interior as they violate the bound (\ref{lim2}).  If we impose (\ref{lim3}), however, all the solutions in the interior will be regular (up to D-brane singularities) and the approximations under good control.    Then combining (\ref{lim3}) and (\ref{Mlims}) for $k \leqq 7$ we will have regular negative mass solutions if
\be \label{flim1}
1.050 \lessapprox \frac{r_1}{r_0} < y_2
\ee
while for $k \geq 8$ the limits of (\ref{lim3}) are stronger than those of (\ref{Mlims}) and so we have good negative mass solutions if
\be \label{flim2}
1.050 \lessapprox \frac{r_1}{r_0} \lessapprox 1.303
\ee
For $k = 5$ the value of $r_1/r_0$ which minimizes $M_0$ (\ref{M0min}) is just allowed by (\ref{flim1}), although for $k \geq 6$ the mass will be minimized for $r_1/r_0$ at the lower bounds of (\ref{flim1}) and (\ref{flim2}).  Note if we take any fixed $r_1/r_0$ in the range of (\ref{flim1}) or  (\ref{flim2}), depending on the value of $k$, we obtain a regular solution whose negative mass scales as the area of the bubble (${r_0}^4$) and may be made arbitrarily negative.

	There is nothing particularly special about the precise form of the family of initial data described above and it turns out there are a variety of examples with similar behavior.  Consider, for example, matching
	$f$ and $g$ as above but allowing a more generic form for $h$; for $r_0 \leq r \leq r_1$
	\be \label{genh}
	h = h_0 (r - r_0) \gamma
	\ee
and then matching h onto some desired function for $r > r_1$.  If one matches h onto a function which rapidly goes from some $h_0$ to the asymptotic value or matches (\ref{genh}) onto some simple polynomials ($R^2 ( 1 - r_0^4/r^4)$ and $R^2 ( 1 - r_0^8/r^8)$ to be precise) one finds qualitatively similar behavior to that above.   In fact, we have examined a variety of other possible initial data and it is reasonably clear that any initial data will have such behavior.   That is, for $k \leq 3$ any regular initial data has positive mass but if $k \geq 4$ there are solutions with negative energy proportional to the area of the bubble and, with the possible exception of $k$, no large numbers are produced.  Of course, this is what one naively expects.

\section{Dynamics}

It is straightforward to check that none of the bubbles we have considered are static.  Despite looking, we have not found any full spacetime solutions with the appropriate boundary conditions aside from the class explored by \cite{HOP}.   Hence the detailed description of the classical evolution of these states must apparently be addressed numerically.    We may, however, make some qualitative observations about the possible dynamics of any bubbles in spacetimes with the desired asymptotics.

Since we have seen a variety of bubbles whose energy becomes more negative as they become larger, dynamically we expect the bubbles in this initial data to tend to expand.   One might have thought the bubbles would expand at nearly the speed of light and could get to timelike infinity.\footnote{By timelike infinity we mean the timelike boundary of AdS.  The reader is free to read ``null infinity'' if he or she prefers.}   However, gravitational radiation and the reflecting boundary conditions we wish to impose on AdS-CFT provide a limit to this expansion.

For any expanding bubble that is regular in any sense, there will be some time-dependent region outside the bubble which will consist of small deviations from the background spacetime and which we may interpret as gravitational radiation.  If a bubble never stopped expanding,  this shell of radiation would bounce back and forth between the bubble and the boundary and becoming increasingly blueshifted as it did so.  Hence, any small amount of radiation would ultimately be blueshifted to the point where the amount of energy in gravitational radiation is unboundedly large.  In physical terms, there is the development of a radiation pressure which acts to confine the bubble.  This pressure becomes arbitrarily strong as a bubble expands to larger and larger radii.  In other words, as usual, AdS is acting like a box which confines excitations to its interior.

To thoroughly explore this possibility, however, let us consider whether there might be bubbles whose negative energy grows sufficiently quickly to overwhelm the energy in radiation.  There are only two ways to achieve such a result--the shell of radiation must become thinner and thinner as the bubble expands out or the bubble's energy must grow increasingly negative at least as fast as the energy in the radiation shell increases.  Let us consider the latter first.  In this case, the amount of energy in gravitational radiation will grow at least as fast as a blueshift factor times ${r_0}^4$, recalling that volumes grow like areas at large distance in AdS.  Despite a reasonably thorough search, we have not found any bubbles whose mass grows much faster than their area. 
Even if they do exist, it is difficult to believe the large number compensating for the blueshift factor would not make the curvature of the bubble correspondingly large and hence produce Planck scale curvatures at some point.   Similarly, in the first scenario at sufficiently large radius the shrinking shell would become of the Planck  thickness and the semiclassical approximation would fail.  In either one of the above scenarios, the frequency of the gravitational radiation grows as the bubble expands and before the bubble could get to infinity one would encounter Planckian frequencies.  Then, in any case, no classical analysis can reliably predict the expansion of a bubble to infinity.

Aside from the above argument, if a bubble ever did get to timelike infinity it would violate the conventional boundary conditions one imposes on $AdS_p \times \mathcal{M}_q$.  Usually one fixes the asymptotic metric, i.e. (\ref{AdSBdy}) and (\ref{SBdy}).  But if the bubble gets to infinity, the asymptotic metric will be changed.  This is simply a matter of definition; if the asymptotic metric is not changed the bubble does not get to infinity.  Note further the presence of a bubble is signaled by a pinching off of a cycle in the compact manifold (in the most interesting case a q-sphere) but there is no such cycle in the asymptotic metric we are fixing.  Even if one defines the asymptotic metric in the AdS directions only up to a conformal factor (see, e.g., \cite{Abhay,HIM} and references therein) the form of the metric in the remaining compact directions still keeps the bubble from getting out to infinity.  In other words, any bubble trying to get out to infinity will ``bounce off''.

It is also worth noting that the bubbles we have discussed include D3-branes and if a bubble got to infinity it would have performed the rather remarkable task of transporting positive rest mass to infinity, perhaps even in finite time.  Even worse, if the bubble hits the boundary one would be forced to conclude matter and charge are flowing through it, violating the reflecting boundary conditions we wish to impose.  Given the above, we must reluctantly conclude that the authors of \cite{HOP} are in fact studying AdS-CFT with unconventional boundary conditions.  The ansatz used there implies the bubbles would get to infinity in finite global time and rather dramatically violate the boundary conditions at that time.  We also must admit our suspicion that the approximations in the analysis of \cite{HOP} must miss some effects of gravitational radiation, as argued above.

While the above argument tells us that bubbles will not get to infinity, it does not tell us what will happen provided we enforce the standard reflecting boundary conditions.   The most likely scenario appears to be that a bubble will classically expand outwards for a time but gravitational radiation eventually halts this expansion and the bubble becomes quasi-static.  While the bubble might be approximately stationary, the spacetime will almost certainly not be due to gravitational radiation.

Alternatively, a bubble might reach a maximum size and then begin to collapse again.  It seems unlikely that such a collapse will proceed to the formation of a singularity.  In addition to violating cosmic censorship in a rather extreme way, the solutions have negative mass and negative masses tend to repel each other.  One would also expect that if the bubbles became very small their behavior should be qualitatively similar to that of the asymptotically flat Kaluza-Klein negative mass bubbles.  In that case, numerical study of collapsing negative mass bubbles has always shown the collapse halts after a time and the bubbles begin expanding \cite{LS}.  In our case, bubbles might well settle down to an approximately static state (plus radiation) after a few oscillations or continue to oscillate indefinitely.   We also mention for completeness the possibility that classical evolution could produce Planck scale curvatures and hence invalidate the semiclassical approximation, as above, although we know of no reason why such outcomes should be expected.

\section{Discussion}

We have shown that supergravity admits solutions which are asymptotically $AdS_5 \times S_5/Z_k$ with arbitrarily negative energy for $k \geq 4$.  These solutions describe bubbles that are regular up to curvature singularities due to smeared D-3 branes.   Since these solutions should be perfectly good in string theory, the AdS-CFT correspondence, at least as usually stated, must include them.  As we have argued, such bubbles may never expand to infinity; indeed there is good reason to believe there is an upper limit to the amount any particular bubble may expand.   This leaves, however, several possibilities for the ``end state'' of the classical evolution of a bubble and it remains an open problem to determine which are realized.

We expect that quantum mechanically these negative energy bubbles will be nucleated rapidly along with enough matter, presumably either radiation or black holes, to ensure that energy is conserved.  One would like to provide definitive evidence for or against this proposition and to bring some qualitative measure to what ``rapid'' means, presuming the word is appropriate.   Note this rate is important for determining the ultimate fate of the bulk spacetime.  While the spacetime is, it appears, entirely unstable, any single bubble will only remove a finite volume of AdS.  Unlike the situation of \cite{HOP} where one relaxes the boundary conditions to the point where a single bubble can consume the entire space, the only way the entire space could be removed in any finite time (for the usual boundary conditions) is for the bubble production rate to diverge.  This is not necessarily inconceivable, for one might conjecture the production rate is roughly democratically spread over the uncountably many negative energy bubbles.

The entire consumption of the spacetime would also require that the nucleation effect is not significantly damped as one goes to larger and larger radii.  In most circumstances, and as we have seen above, AdS tends to confine nontrivial states to its interior.  On the other hand, one might argue that the states with larger and larger bubbles, and the subsequently large number of matter degrees of freedom, should be favored entropically and quantum mechanically.   If the production rate did diverge throughout AdS, it would mean that there is no region in which the bulk spacetime is semiclassical and in particular would require some definition beyond the usual one of asymptotically $AdS_p \times S_q$ boundary conditions.

Alternatively, bubbles might be nucleated throughout AdS but at a finite rate so that  any given region of the spacetime will eventually be consumed by a bubble but there will always be regions of spacetime remaining.   The nucleation effect also might be substantially confined to the interior of AdS: quickly eliminating regions smaller than an AdS scale but leaving the asymptotics intact.  At the moment we do not have tools to decide among these possibilities.

Given the above state of affairs for the bulk, one would like to understand the implications for AdS-CFT.  The status of the orbifolded gauge theory at strong coupling is largely an open problem.  The Hamiltonian of the original $\mathcal{N} = 4$ SYM theory is bounded below as a result of supersymmetry but the orbifold we have chosen eliminates this simple argument.  One might have been inclined to view the orbifold on the gauge theory as a rather mild operation that should not make the theory unstable.  There would seem to be significant tension between such a view and the AdS-CFT correspondence.   Besides the correspondence, there do not seem to be any tools at hand to study the strongly coupled theory or even to determine whether the Hamiltonian is bounded from below.

The orbifolded gauge theory can and has been studied at weak t'Hooft coupling \cite{Dymaskyetal}.  It has some unusual features, perhaps most notably that the couplings of twisted sector operators run and break conformal invariance.  The theory requires an ultraviolet cutoff for these couplings, although whether this is a fundamental problem or, like the same feature in $\lambda \phi^4$, indicates the emergence of new phenomena at a given energy scale is not yet clear.   However, there are no tools like supersymmetry or large conserved charges available to allow us to continue the results from weak to strong coupling.  In fact, the AdS-CFT correspondence (presuming the spacetime is sufficiently stable that a semiclassical region exists) indicates the breaking of conformal invariance in the twisted sectors disappears at large t'Hooft coupling; in particular the twisted-trace two-point function  is conformal \cite{AdamsSilverstein}.  So in this sense at least the strongly coupled theory seems to be better behaved.  On the other hand, there are several examples of nonsupersymmetric field theories that are believed to make sense at weak, but not strong, coupling \cite{APS, Klebanov}.

Given the existence of negative energy states on the gravitational side, there seem to be three possibilities for the status of AdS-CFT in this context: the field theory is well defined and stable but the correspondence fails in this circumstance, the correspondence holds and the field theory is unstable, or the correspondence is valid and the theory somehow stable despite the fact the Hamiltonian is unbounded from below.  The last seems difficult to believe, especially in a theory at strong coupling,  and the apparent instability of the bulk.  However, we are not aware of a way to definitively rule out this possibility.  Note that if the field theory is unstable, one might hope to describe the decay of a state with given energy.  We have described a variety of possible decay mechanisms in the bulk -- if AdS-CFT continues to hold in this situation the description of the decay must match that of the gauge theory.  On the other hand, if both sides are hopelessly pathological it becomes increasingly important to understand why this apparently mild orbifolding is so dangerous and the limits of modifying the AdS-CFT correspondence.

\vskip 1cm
\centerline{\bf Acknowledgments}
\vskip .5cm

We would like to thank R. Myers, F. Cachazo and A. Buchel for comments and G. T. Horowitz, J. Polchinski, and D. Marolf for significant correspondence.   This work was supported by the Natural
Sciences and Engineering Research Council of Canada and completed in part at the Perimeter Institute, for whose hospitality we express our appreciation.

 \end{document}